\documentclass[pra,reprint,amsmath,amssymb,aps,superscriptaddress]{revtex4-2}

\usepackage{amsthm}
\usepackage{algorithm}
\usepackage[noend]{algpseudocode}

\usepackage{tabularx}
\usepackage{graphicx}
\usepackage{dcolumn}
\usepackage{bm}

\usepackage{appendix}
\usepackage{comment}

\usepackage[shortlabels]{enumitem}

\usepackage[colorlinks, linkcolor=blue,anchorcolor=blue,citecolor=blue,urlcolor=blue]{hyperref}

\usepackage{soul}

\begin{document}

\title{Efficient algorithms to solve atom reconfiguration problems. III.~The bird and batching algorithms and other parallel implementations on GPUs}

\author{Fouad~Afiouni}
\affiliation{Department of Computer Science, American University of Beirut, Lebanon.}
\affiliation{School of Electrical and Computer Engineering, Purdue University, USA.}
\author{Remy~El~Sabeh}
\affiliation{David R. Cheriton School of Computer Science, University of Waterloo, Canada.}
\author{Naomi Nishimura}
\affiliation{David R. Cheriton School of Computer Science, University of Waterloo, Canada.}
\author{Izzat~El~Hajj}
\affiliation{Department of Computer Science, American University of Beirut, Lebanon.}
\author{Amer~E.~Mouawad}
\affiliation{Department of Computer Science, American University of Beirut, Lebanon.}

\affiliation{David R. Cheriton School of Computer Science, University of Waterloo, Canada.}
\author{Alexandre~Cooper}
\email[]{alexandre.cooper@uwaterloo.ca}
\affiliation{Institute for Quantum Computing, University of Waterloo, Canada.}
\affiliation{Department of Physics and Astronomy, University of Waterloo, Canada.}

\begin{abstract}
We present efficient implementations of atom reconfiguration algorithms for both CPUs and GPUs, along with a batching routine to merge displacement operations for parallel execution. Leveraging graph-theoretic methods, our approach derives improved algorithms that achieve reduced time complexity and faster operational running times. First, we introduce an enhanced version of the redistribution-reconfiguration (red-rec) algorithm, which offers superior operational and runtime performance. We detail its efficient implementation on a GPU using a parallel approach. Next, we present an optimized version of the assignment-reconfiguration-ordering (aro) algorithm, specifically tailored for unweighted grid graphs. Finally, we introduce the bird algorithm to solve reconfiguration problems on grids, achieving performance gains over both red-rec and aro. These algorithms can be used to prepare defect-free configurations of neutral atoms in arrays of optical traps, serve as subroutines in more complex algorithms, or cross-benchmark the operational and runtime performance of new algorithms. They are suitable for realizing quantum circuits incorporating displacement operations and are optimized for real-time operation on increasingly large system sizes.
\end{abstract} 

\maketitle 

\section{Introduction}
Configurations of individual neutral atoms in arrays of optical traps provide a versatile platform for quantum information processing~\cite{Browaeys2020, Morgado2021}. These atoms can be displaced in space to deterministically prepare defect-free configurations of atoms with specific geometries~\cite{Endres2016, Barredo2016, Barredo2018} and continuously replenish atoms from a separate reservoir~\cite{Pause2023, Norcia2024, Gyger2024}. Moreover, these atoms encode quantum information in their internal degrees of freedom, which are effectively decoupled from their external degrees of freedom over relevant time scales. This decoupling allows atoms to be spatially displaced with minimal loss while maintaining coherence among internal states, thereby enabling the creation of entangled gates between distant atoms~\cite{Bluvstein2022}, the partitioning of computational tasks into separate spatial regions~\cite{Bluvstein2024}, and the efficient realization of algorithms and protocols requiring non-local spatial connectivity~\cite{Bravyi2024, Maskara2025}.

Finding the set of displacement operations to transform an arbitrary initial configuration into a target configuration requires solving atom reconfiguration problems using efficient atom reconfiguration algorithms~\cite{Cooper2024, Schymik2020}. These algorithms seek to minimize metrics such as the number of displaced atoms and the total displacement distance, both of which correlate with atom loss. These algorithms can be further improved using batching routines, which seek to perform displacement operations in parallel while satisfying hardware constraints~\cite{Tian2023, Wang2023}.

Various reconfiguration algorithms, based on exact, approximation, and heuristic methods, have been developed and experimentally realized~\cite{OhldeMello2019, Schymik2020, Ebadi2021, Tao2022, Tian2023, Cimring2023, ElSabeh2023, Wang2023}. While these algorithms have been primarily designed with operational considerations in mind, there remains a need for further development of efficient algorithms derived from first principles, such as those based on graph theory. Establishing formal theoretical foundations would help identify bounds on reconfiguration performance while enabling the evaluation of different algorithms and their hardware implementations.

There is also a growing need for formal results in two main areas. First, efficient algorithms and their implementations on typical processors, such as central processing units (CPUs) and graphics processing units (GPUs), are required. This includes improving the time complexity of algorithms by restricting the search to specific problem instances. Second, batching routines are needed to combine displacement operations for parallel execution, ideally in a way that is decoupled from the atom reconfiguration algorithms.

\begin{table}
\caption{Comparison of reconfiguration algorithms, specifying the geometry of their input graph of size $n$, their target atom configuration, and their improved time complexities.
}
\label{table:outline}
\begin{ruledtabular}
\begin{tabular}{llll}
Algorithm & Graph & Configuration & Complexity \\ \hline
exact 1D & Paths/Chains & Arbitrary & $\mathcal{O}(n^2)$ \\
red-rec~\cite{Cimring2023} & Grids & Arbitrary & $\mathcal{O}(n\sqrt{n})$ \\
aro~\cite{ElSabeh2023} & Grids & Arbitrary & $\mathcal{O}(n^4)$ \\
bird & Grids & Center-compact & $\mathcal{O}(n\sqrt{n})$ \\
\end{tabular}
\end{ruledtabular}
\end{table}

In this paper, we establish formal results on atom reconfiguration algorithms~(see Table~\ref{table:outline}), quantify their operational and runtime performance on a low-latency reconfiguration system, and introduce a batching routine independent of the reconfiguration algorithm. Our approach bridges the gap between two disparate fields: experimental quantum sciences and graph theory, specifically combinatorial reconfiguration problems and combinatorial optimization on graphs. We leverage graph-theoretic methods and provide formal proofs of convergence and complexity.

We improve upon three previously introduced algorithms: the exact 1D algorithm for paths/chains, the redistribution-reconfiguration (red-rec) algorithm for grid-embedded geometries~\cite{Cimring2023}, and the assignment-reconfiguration-ordering (aro) algorithm for arbitrary graphs~\cite{ElSabeh2023}. We study efficient implementations of red-rec using parallel execution on GPUs~(Sec.~\ref{sec-red-rec-parallel}). We then introduce the bird algorithm as a generalization of red-rec, specifically designed for grids~(Sec.~\ref{sec-bird}). We finally present a batching routine that optimizes the parallel execution of displacement operations~(Sec.~\ref{sec-batching}), and quantify performance~(Sec.~\ref{sec-operational-benchmarking}). These results may serve as subroutines or benchmarks for existing and future algorithms.

\section{Atom Reconfiguration Problems}
We start by reviewing the key concepts and formal definitions related to atom reconfiguration problems from both operational~(Sec.~\ref{subsec-operational}) and graph-theoretic~(Sec.~\ref{subsec-graph-theoretical}) perspectives. We then formulate the atom reconfiguration problem as an optimization problem~(Sec.~\ref{sec-matching-flows-systems}), where the objective is to find a solution that minimizes the distance along a path system. Familiar readers may directly jump to reconfiguration problems~(Sec.~\ref{sec-algo-impl}), the batching routine~(Sec.~\ref{sec-batching}), or numerical cross-benchmarking results~(Sec.~\ref{sec-operational-benchmarking}).

\subsection{Atom reconfiguration problems in practice}\label{subsec-operational}
Atom reconfiguration problems~\cite{Schymik2020, Cooper2024, Cimring2023} seek to produce a control protocol that transforms a given arbitrary configuration $\mathcal{C}_0$ of $N_a^0$ atoms into a given target configuration $\mathcal{C}_T$ of $N_a^T \leq N_a^0$ atoms.
A configuration of atoms is contained in an array of optical traps, $\mathcal{A}(V)$, defined by its spatial arrangement or \emph{geometry}, $V = \{\vec{v}_j~|~\vec{v}_j = (v_{j_x}, v_{j_y}) \in \mathbb{R}^2, 1 \leq j \leq N_t\}$, where $N_t$ is the number of traps in the optical trap array. We focus on square lattices (grids) of $N_t=N_{t_x}\times N_{t_y}$ traps in the plane where $v_{j_x}=x_0+j_x\delta_x$ and $v_{j_y}=y_0+j_y\delta_y$ for $0\leq j_x\leq N_{t_x} - 1$, $0 \leq j_y \leq N_{t_y}-1$, $(x_0, y_0)$ is the origin of the array, and $\delta_x$ and $\delta_y$ are the grid spacing constants.

The solution to an atom reconfiguration problem is a control protocol. The control protocol consists of a sequence of extraction-displacement-implantation (EDI) cycles that each extract, displace, and implant a single atom from one static trap to another using a secondary array of dynamic traps. Each EDI cycle executes a sequence of elementary control operations that include elementary transfer operations and elementary displacement operations. A transfer operation extracts (implants) an atom from (into) a static trap into (from) a dynamic trap. An elementary displacement operation  displaces a dynamic trap containing an atom from one static trap to another by an elementary displacement step $\delta x$ or $\delta y$. Simply put, and given that we focus on grids, an elementary displacement step consists of moving a dynamic trap containing an atom from some grid position to one of the (at most) four neighboring grid positions.  

Given an initial and a target configuration of atoms, the reconfiguration problem 
is solved, the control protocol is executed, and a measurement is performed to check whether or not the updated configuration of atoms contains the target configuration. 
In the presence of loss, the reconfiguration problem might have to be solved multiple times through multiple reconfiguration cycles until the target configuration is reached (success) or is no longer reachable (failure); failure occurs when more than $N_a^0 - N_a^T$ atoms are lost. 

\subsection{Atom reconfiguration problems on graphs}\label{subsec-graph-theoretical}
Atom reconfiguration problems can be viewed as reconfiguration problems on graphs~\cite{Cooper2024,Cimring2023,ElSabeh2023}. 
A configuration of indistinguishable atoms contained in an array of optical traps is represented as a collection of tokens placed on a subset of the vertices of a graph, $G$, where $V(G)$ and $E(G)$ denote the vertex set and edge set of $G$, respectively, with $|V(G)| = n$ and $|E(G)| = m$. We assume that each graph is finite, simple, connected, and undirected (see Ref.~\cite{diestel} for standard graph terminology). For weighted graphs, we use $w_G: E(G) \rightarrow \mathbb{N}^{+}$ to denote the edge-weight function, where $w_G(e)$ is positive for all $e = \{u,v\} = uv \in E(G)$. In the following, we omit subscripts when the context is clear.

A \emph{path} in a graph $G$ is a walk whose sequence of vertices comprises distinct vertices. A \emph{walk} (of length $\ell$) in $G$ is a sequence of vertices in $V(G)$, $(v_0, \ldots, v_\ell)$, such that $v_iv_{i + 1} \in E(G)$ for all $i \in \{0, \ldots, \ell-1\}$, where $\{v_1, v_2, \ldots, v_{\ell-1}\}$ are the \emph{internal} vertices of the walk. We define a \emph{cycle} in $G$ as a walk of length $\ell \geq 3$ that starts and ends at the same vertex, $v_0 = v_\ell$, and whose internal vertices form a path. 
The weight of a path $P$ in $G$ is $w_G(P) = \sum_{e \in P}{w_G(e)}$. When the graph is unweighted, or all of its edges have the same weight, in which case the graph is said to be \emph{uniformly weighted}, then the weight of $P$ is equal to the number of edges in $P$ multiplied by the weight; the weight is assumed to be $1$ for unweighted graphs. 
For $u,v \in V(G)$, let $P_{uv}$ denote a minimum-weight path from $u$ to $v$. We say $P_{uv}$ is a \emph{shortest path} from $u$ to $v$ and the \emph{distance} from $u$ to $v$, denoted by $d_G(u,v)$, is equal to the weight of $P_{uv}$, i.e., $d_G(u,v) = w_G(P_{uv})$. For unweighted graphs, the distance from $u$ to $v$ corresponds to the minimum number of edges required to get from $u$ to $v$ (by convention, we set the distance to infinity when $u$ and $v$ belong to different components). 

Even though some of our algorithms, e.g., the aro algorithm~\cite{ElSabeh2023}, were designed for general (positive) edge-weighted graphs, here we mostly focus on (unweighted) \emph{path/chain graphs} and (unweighted) \emph{grid graphs} to achieve the claimed improvements in running time. An \emph{$n$-path graph} is a graph on $n$ vertices with vertex set $\{v_i \mid 0 \leq i \leq n - 1\}$ and edge set $\{v_iv_{i+1} \mid 0 \leq i \leq n - 2\}$. A \emph{$(W \times H)$-grid graph} is a graph on $n = W\times H$ vertices with vertex set $\{(x, y) \mid 0 \leq x \leq W - 1,~0 \leq y \leq H - 1\}$, for $W,H\in\mathbb{N}^{+}$. Two vertices $v=(x, y)$ and $v'=(x', y')$ ($v \neq v'$) are adjacent in a grid graph, and thus connected by an edge, if and only if $|x - x'| + |y - y'| \leq 1$. We use $W$ to denote the \emph{width} of a grid graph, which we also refer to as the number of \emph{columns}. Similarly, we use $H$ to denote the \emph{height} of a grid graph, which we also refer to as the number of \emph{rows}. 

In addition to the graph $G$, the atom reconfiguration problem requires defining the initial (source) and desired (target) configurations of atoms. The traps containing the atoms in the source and target configurations are identified as subsets of vertices $S \subseteq V(G)$ and $T \subseteq V(G)$, respectively. We assume that $|S| \geq |T|$ since, otherwise, the problem does not admit a solution. Note that $S$ and $T$ need not be disjoint. Each vertex in $S$ has a token on it. We aim to select a subset $S^\star$ of $S$ and move the tokens in $S^\star$ along the edges of the graph so that all vertices in $T$ eventually contain tokens, with the constraint that each vertex in the graph contains at most one token after every elementary displacement operation. The selection of $S^\star$ and the selection of the edges along which the tokens are to be moved are done in a way that minimizes a mixed objective function (see Sec.~\ref{sec-matching-flows-systems}). 

Here, a \emph{move} of token $\tau_i$ ($1 \leq i \leq |S^\star|$) from vertex $u$ to vertex $v$, which is equivalent to a sequence of elementary displacement operations, is \emph{unobstructed} whenever $\tau_i$ is on $u$ and the selected path $P$ (discussed formally in the next section) from $u$ to $v$ in $G$ associated with $\tau_i$ is free of tokens (except for $\tau_i$); otherwise, we say that the move is \emph{obstructed} and call each token $\tau_j$ ($j \neq i$) on $P$ an \emph{obstructing token}.
If we attempt to move a token along a path that is not free of tokens, then we say that this move causes a \emph{collision}. Because a collision induces the loss of the colliding atoms, moves that cause collisions are replaced by sequences of moves that do not cause collisions.
Indeed, if the move of token $\tau_i$ from $u$ to $v$ along $P$ is obstructed, then, assuming $v$ is not occupied by a token, we can always replace the move of $\tau_i$ by a sequence of unobstructed moves involving the obstructing tokens~\cite{ElSabeh2023}.
A solution to an atom reconfiguration problem is thus a (partially-ordered) sequence of moves to be executed in order, such that each move is unobstructed at the time of its execution, and all vertices in $T$ contain tokens after all moves are executed. 

\subsection{Solving atom reconfiguration problems}\label{sec-matching-flows-systems}
Two natural metrics to minimize while solving atom reconfiguration problems are the total number of elementary displacement, which is equivalent to the total distance traversed by atoms, and the total number of displaced atoms. The latter provides a lower bound on twice the number of transfer operations, since each displaced atom must be extracted and implanted at least once. Intuitively, minimizing these metrics translates into a smaller probability of loss and, consequently, a higher probability of success. Computing solutions that minimize both metrics simultaneously cannot be done efficiently due to the intrinsic computational complexity of the problem~\cite{ElSabeh2023}. Moreover, computing a solution that minimizes the number of displaced atoms is an \textsf{NP}-complete problem~\cite{Calinescu2007}, and is therefore unlikely to admit efficient, i.e., polynomial-time, (exact) algorithms. This statement remains true even when instances are restricted to unweighted grid graphs~\cite{Calinescu2007}.

Fortunately, finding solutions that minimize the total number of displacements is possible in polynomial time. Solving this problem requires first pairing/matching/assigning each vertex $v_{t} \in T$ with a distinct vertex $v_{s} \in S$, such that the sum of distances between the vertices of each pair is minimized. In other words, each target vertex in $T$ must be paired with a distinct source vertex from $S$, with the goal of minimizing the sum of distances.

There are several possible approaches for computing such a pairing of vertices; we now discuss those approaches that are relevant to atom reconfiguration problems on grids. Starting with a graph $G$ and $S,T \subseteq V(G)$, we construct an edge-weighted \emph{bipartite} graph $H$. This bipartite graph has a vertex set that is partitioned into two sets such that no edge connects two vertices within the same set. We denote this bipartition by $(L = S, R = T)$. For every vertex $v_s \in L$ and every vertex $v_t \in R$, we add the edge $v_sv_t$ to $E(H)$ and we let $w_H(v_sv_t) = d_G(v_s,v_t)$, i.e., the weight of the edge in $H$ is equal to the distance between $v_s$ and $v_t$ in $G$ (with $v_s$ possibly equal to $v_t$ in which case the distance/weight is zero). Given a set of edges $M \subseteq E(H)$, we say $M$ is a \emph{matching} whenever no vertex of $H$ belongs to more than one edge in $M$. We use $V(M)$ to denote the vertices belonging to edges of $M$ and we say that $M$ \emph{saturates} a vertex-subset $Q$ whenever $Q \subseteq V(M)$. For edge-weighted graphs, we define $w(M) = \sum_{e \in M}{w(e)}$ and we say $M$ is a \emph{minimum-weight (or distance-minimizing) matching} (saturating $Q$) whenever there exists no matching $M'$ in $H$ (saturating $Q$) with $w(M') < w(M)$. Clearly, we require a matching to saturate a specific vertex subset (in our case, $T$) and whenever we mention a minimum-weight matching we assume that $T$ is saturated. It can be shown that computing the required pairing between vertices in $S$ and vertices in $T$ is equivalent to finding a minimum-weight matching $M$ in $H$ that saturates $T$. Several polynomial-time algorithms are known for computing such a matching, e.g., the Hungarian algorithm~\cite{edmonds1972theoretical} and the minimum-cost maximum-flow (MCMF) algorithm~\cite{ford1956maximal}. More efficient algorithms are known for special cases, such as when $H$ is constructed from a path graph $G$~\cite{DBLP:journals/dm/KarpL75}. We provide a more detailed discussion of these algorithms in Sec.~\ref{sec-algo-impl} 

Clearly, it is not enough to just maintain a distance-minimizing matching; as such a matching does not provide the information needed for generating the elementary displacement operations of the tokens, i.e., the unobstructed displacement trajectories. We resolve the issue by maintaining an arbitrary shortest path $P_i$ (in $G$) for every matched pair, i.e., for every $v_{s_i}v_{t_i} \in M$. We call this set of paths a path system, which is computed using either several breadth-first search (BFS) calls or using an all-pairs shortest-path (APSP) algorithm. 
In our previous work, we introduced path systems without any restrictions on the lengths or weights of the paths~\cite{ElSabeh2023}. However, for this work, restricting path systems to shortest paths is sufficient and simplifies the presentation. Formally, we define a \emph{path system} $\mathcal{P}$ in $G$ as a set of shortest paths, $\mathcal{P} = \{P_1, P_2, \ldots, P_k\}$, in which each path $P_i \in \mathcal{P}$ for $i \in \{1, \ldots, k\}$ is a shortest path from $v_{s_i}$ (source vertex) to $v_{t_i}$ (target vertex), which we denote by $\{v_{s_i}, v_1, v_2, \ldots, v_{t_i}\}$. Single-vertex paths with $v_{s_i} = v_{t_i}$ are also allowed. The weight of a path system $\mathcal{P}$ in $G$ is given by the sum of the weights of its paths, $w(\mathcal{P}) = \sum_{P \in \mathcal{P}}{w(P)} = \sum_{P \in \mathcal{P}}{\sum_{e \in P}{w(e)}}$. 

We say that a move of token $\tau_i$ associated with path $P_i \in \mathcal{P}$ is \emph{executable} whenever the target vertex $v_{t_i}$ does not contain a token;
an unobstructed move is trivially executable, whereas an obstructed move can always be converted to a sequence of unobstructed moves, assuming $v_{t_i}$ contains no token. A path system $\mathcal{P}$ is said to be \emph{valid (for $T$)} or \emph{$T$-valid} whenever there exists some ordering of the moves/paths that makes them executable, and executing all the moves associated with $\mathcal{P}$ results in each vertex in $T$ having a token on it (we assume that the weight of a path system is infinity whenever it is not valid). Clearly, in a valid path system, all source vertices are distinct and all target vertices are distinct, although some source vertices can be the same as some target vertices. 
We note that for any valid path system, we can always find an executable move, unless the problem has already been solved with all vertices in $T$ containing tokens. 
We also note that, before any move is executed, whenever we have a token $\tau$ on some target or internal vertex of a path $P \in \mathcal{P}$, then there must exist another path $P' \in \mathcal{P}$, $P' \neq P$, for which the token $\tau$ is on the source vertex. Using the previous observations, it becomes easy to show that our matching procedure (in the bipartite graph $H$) combined with shortest paths computation (in $G$) as well as conversion of obstructed moves (if needed) produces a valid (partially-ordered) path system~\cite{ElSabeh2023}. 

Putting it all together, a solution to an atom reconfiguration problem can now be viewed as a (partially-ordered) path system $\mathcal{P}$ such that executing the moves associated with $\mathcal{P}$ (respecting the partial order) places a token on every vertex in $T$ while avoiding collisions. Given that we are dealing with unweighted grid graphs, each path  $P \in \mathcal{P}$ having $|E(P)|$ edges corresponds to exactly $|E(P)|$ elementary displacement operations.  

\section{Atom Reconfiguration Algorithms}\label{sec-algo-impl}

In this section, we provide a detailed presentation of several proposed or improved atom reconfiguration algorithms, along with their implementation details. 
In particular, we propose an improved 1D reconfiguration algorithm (Sec.~\ref{sec-exact_1d}) and its parallel implementation (Sec.~\ref{sec-exact_1d-parallel}). We also describe improvements to the previously proposed red-rec algorithm (Sec.~\ref{sec-red-rec}) and its parallel implementation (Sec.~\ref{sec-red-rec-parallel}), introduce a newly proposed reconfiguration algorithm for grid graphs with centered targets, which we call the bird algorithm (Sec.~\ref{sec-bird}), and present improvements to the previously proposed aro algorithm (Sec.~\ref{sec-aro}).

\subsection{The exact 1D algorithm (serial implementation)}\label{sec-exact_1d}

One of the core subroutines of the red-rec and bird algorithms is a 1D reconfiguration subroutine, known as the exact 1D algorithm. This subroutine reconfigures individual columns of a grid through two key operations: (1) assigning source tokens to target traps and (2) generating moves along unique shortest paths to realize the target configuration.

In the original red-rec algorithm, these two operations were combined. In this work, we separate them for modularity and enhance each step to improve the efficiency and versatility of our system. For the assignment routine, we use a simpler algorithm than the one used for the original red-rec implementation and develop a generalized assignment routine required by the bird algorithm. For move generation, we introduce an improved subroutine that computes a partial ordering of moves, which is then used by the batching routine (see Sec.~\ref{sec-batching}).

\subparagraph{Assignment.} The assignment algorithm computes a distance-minimizing matching between $S$ and $T$ that saturates $T$ in linear time by exploiting the simple structure of path graphs to avoid having to construct the bipartite graph $H$ (discussed in Section~\ref{sec-matching-flows-systems}).
When $|S| = |T|$, one can easily compute (in linear time) a distance-minimizing matching by two passes over the graph; once left-to-right and once right-to-left, where we assume that the vertices of the path are ordered horizontally from left to right.
The challenge occurs 
when $|S| > |T|$~\cite{DBLP:journals/dm/KarpL75,DBLP:journals/jal/AggarwalBKKS95}. 
The prior red-rec algorithm relied on the linear assignment algorithm of Karp and Li~\cite{DBLP:journals/dm/KarpL75}.
In this work, we instead use a simpler version of the algorithm proposed by Aggarwal et al.~\cite{DBLP:journals/jal/AggarwalBKKS95}.

\subparagraph{Generalized assignment.} While the red-rec algorithm requires an assignment subroutine where the number of tokens mapped to a vertex in $S \cup T$ is one, our proposed bird algorithm requires a more general subroutine that assumes that the vertices in $S$ can have multiple tokens mapped to them (see Sec.~\ref{sec-bird} for details).
To support the bird algorithm, we implement both the simple variant as well as the generalized variant with multiple tokens mapped to a source vertex.
This mapping is only a virtual one to solve particular cases in 2D.
We never require multiple atoms to physically occupy the same trap in practice.


\subparagraph{Generating moves.}
The 1D reconfiguration subroutine in the previous red-rec algorithm generates a sequential list of batched or unbatched moves; however, the batching routine requires a partial ordering of moves to properly group them into batches.
We construct this partial ordering as follows.
After constructing a distance-minimizing matching $M$ (via assignment), we generate a path system $\mathcal{P}$ containing the unique shortest path between the vertices of every matched pair.
Next, we label every path in $\mathcal{P}$ as either
    \emph{right-oriented} (the source vertex appears to the left of the target vertex in $G$),
    \emph{left-oriented} (the target vertex appears to the left of the source vertex in $G$), or 
    \emph{isolated} (the source vertex is equal to the target vertex).
Given that $M$ is distance minimizing, we know that no vertex of $G$ can appear in both a left-oriented and a right-oriented path.
Moreover, no two paths in $\mathcal{P}$ can share the same source or target vertex.
As our final goal is to produce a partially-ordered list of elementary displacement operations, we need to address one remaining issue.
This issue occurs when a left-oriented (or right-oriented) path $P'$ is fully contained within another left-oriented (or right-oriented) path $P$, meaning all vertices of $P'$ are internal vertices of $P$ (this occurs since the assignment algorithm prefers to match vertices that are closer to each other). 
We resolve this potential collision issue by modifying the path system; for any such problematic pair of paths, $P$ and $P'$, we swap their target vertices.
We then fix the path system in linear time by traversing the graph $G$ once from left to right (fixing right-oriented paths) and once from right to left (fixing left-oriented paths).
Finally, we compute a partially-ordered list of elementary displacement operations such that the in-order execution of the operations in the list solves the reconfiguration problem without any obstructions.
The existence of such a list is guaranteed by the fact that our path system minimizes total distance, left-oriented paths and right-oriented paths do not share any vertices, and no path is fully contained inside another.
One such partially-ordered list of elementary displacement operations can be obtained by solving right-oriented paths from the rightmost path to the leftmost path, i.e., sort the right-oriented paths by the index of their target vertices in ascending order and add to the list the elementary displacements of each path in descending order.
The same procedure is repeated for left-oriented paths to complete the list.
Any isolated path not addressed in either of the two previous steps is ignored, as the token does not need to move.

\subparagraph{Complexity.}

Our exact 1D algorithm is efficient at solving atom reconfiguration problems on path graphs, also known as chain graphs.
Even though one can compute a matching/assignment in time $\mathcal{O}(n)$, the exact 1D algorithm has a running time of $\mathcal{O}(n^2)$ in the worst case, for an $n$-path graph~$G$.
Asymptotically, the quadratic bound on running time is the best possible as there exist instances that require $\mathcal{O}(n^2)$ elementary displacement operations to be solved, e.g., a path on $2n$ vertices where $S$ is the first $n$ vertices and $T$ is the last $n$ vertices.
This quadratic bound is rarely reached in practice since most instances of the atom reconfiguration problem on paths require a linear number of displacement operations; as atoms are initially distributed with equal probability over the path.
This expectation is confirmed by experimental results showing that our implementation of the exact 1D algorithm scales linearly with input size. 

When source and target vertices can hold more than one token, the assignment problem on paths is solvable in time quadratic in $|S| + |T|$, as noted by Karp and Li~\cite{DBLP:journals/dm/KarpL75}.
However, if we restrict target vertices to single tokens and only allow multiple tokens per source vertex, we can again solve the assignment problem in linear, $\mathcal{O}(n)$, time.
Hence, the generalized exact 1D algorithm also runs in $\mathcal{O}(n^2)$-time in the worst case for an $n$-path graph~$G$.

\subsection{The exact 1D algorithm (parallel implementation)\label{sec-exact_1d-parallel}}

In this section, we describe how we parallelize the exact 1D algorithm, so that it can be used as a subroutine in our parallel implementation of the red-rec algorithm described in Section~\ref{sec-red-rec-parallel}.
To do so, we aim to divide the problem into many independent sub-problems and solve the sub-problems in parallel.
To find independent sub-problems, we look for pairs of vertices that have the same number of source and target vertices between them.

Formally, in a single dimension, our graph $G$ is a path on $n$ vertices, $\{v_0, \ldots, v_{n-1}\}$.
We let $G[i,j]$ denote the subgraph of $G$ consisting of the path from vertex $v_i$ (inclusive) to vertex $v_j$ (inclusive). 
We let $\Delta_i$ denote the difference between the number of sources and the number of targets in $G[0, i]$, i.e., $\Delta_i = |V(G[0,i]) \cap S| - |V(G[0,i-1]) \cap T|$.
We say vertex $v_i$ is at level $\Delta_i$.
We observe that if there exists a source vertex $v_i \in S$ and a target vertex $v_j \in T$ such that $v_i$ and $v_j$ are non-adjacent  and $\Delta_i = \Delta_j$, then the number of source vertices in $G[i+1,j-1]$ is equal to the number of target vertices (we assume without loss of generality that $i < j$ as we can otherwise consider the graph $G[j+1,i-1]$).
In other words, the sub-problem in between $v_i$ and $v_j$, i.e., $G[i+1,j-1]$, is ``self-contained''.

To find these independent sub-problems, we start by computing $\Delta_i$ for each index $i$, $0 \leq i \leq n -1$.
Recall that $\Delta_i$ is the difference between the number of sources and the number of targets in $G[0, i]$.
Hence, we compute $\Delta_i$ for each vertex by performing a parallel prefix sum operation~\cite{wen2022programming}.

Next, the thread associated with each target vertex searches for the two closest source vertices that belong to the same level.
Specifically, for each target vertex $v_q$, a thread searches for $v_p$ and $v_r$ such that $p < q < r$, $\Delta_p = \Delta_q = \Delta_r$, and there exists no source vertex in between $v_p$ and $v_r$ that is at the same level as $v_q$.
We avoid a standard linear search by making use of the previously computed vertex levels to leap over sections of the path.
For example, if a thread looking for a vertex at level \textit{s} has reached a vertex whose level is \textit{s-k}, then it can leap over \textit{k} vertices because the vertex at level \textit{s} must be at least \textit{k} vertices away.

Once each thread has found the independent sub-problems associated with its target vertex, the following step is to guarantee that the sub-problems do not interact, i.e., no tokens have to move between the corresponding sub-paths.
This step can be done based on the computation of \emph{profits} as described in the algorithm of Karp and Li~\cite{DBLP:journals/dm/KarpL75}.
Once the independent and non-interacting sub-problems have been found, they are all solved in parallel.
Finally, each thread associated with a source vertex indicates (by setting a bit in global memory) whether the corresponding token/atom is going to move.

\subsection{The red-rec algorithm (serial implementation)}\label{sec-red-rec}

The redistribution-reconfiguration (red-rec) algorithm is a heuristic algorithm that solves atom reconfiguration problems on grids~\cite{Cimring2023}.
In this paper, we propose an improved version of the red-rec algorithm that leverages the improved 1D reconfiguration algorithm, a new redistribution strategy, and delayed moves.
Before describing our improved algorithm, we first review the original algorithm and its implementation. 








\subparagraph{The original red-rec algorithm.}
The original red-rec algorithm~\cite{Cimring2023} runs on a grid graph $G$, where we assume $n = W \times H = \sqrt{n} \times \sqrt{n}$. Since $G$ is planar, we assume a grid-like planar embedding of the graph in our notation and discussion. We use $\mathcal{C}_i$ and $\mathcal{R}_j$ to denote the graphs induced on the vertices in the $i$-th column and $j$-th row of $G$, respectively. In the planar embedding, we assume that the rows are indexed from bottom to top (starting from $0$) and the columns are indexed from left to right (also starting from $0$). The vertices of a column (or a row) induce a path. We are also given a set $S \subseteq V(G)$ of source vertices, each occupied by a token, and a set $T \subseteq V(G)$ of target vertices, such that $|S| \geq |T|$. We assume that the target configuration $T$ is always \emph{centered} and \emph{compact}, i.e., the target configuration is a $W \times H'$-subgrid centered inside the input grid (in the planar embedding of the input grid), where $H' < H$. We call the subgrid the \emph{target region} and we call the remaining parts (above/below the target region) the \emph{(top/bottom) reservoir region}.
We note that the red-rec algorithm can be adapted to handle arbitrary target configurations, but we omit the details here, as the required changes are highly non-trivial and provide very little additional insights. 

The first step of the original red-rec algorithm computes the \emph{surplus} of each column, where the surplus can be either negative or positive.
In particular, for each column $\mathcal{C}_i$, $0 \leq i < W$, we have surplus equal to $\sigma_i = |S \cap V(\mathcal{C}_i)| - |T \cap V(\mathcal{C}_i)|$.
The elementary displacement operations required to solve every column with a surplus of zero are computed using the exact 1D algorithm (Sec.~\ref{sec-exact_1d}).
These operations are executed to modify the current input instance and subsequently added to the output, i.e., the input instance is modified into an instance where all said columns are solved and the associated moves are appended to the solution (to the partial order of output moves).
We refer to these three sequential steps as \emph{solving} a column.
For a column with a negative surplus, we refer to the surplus as the \emph{distance to saturation}.
When the surplus changes from negative to zero, we say the column becomes saturated.
Therefore, every column must be saturated; otherwise, the problem cannot be solved. 

Once there are no more unsolved columns with a surplus equal to zero, the algorithm enters the main loop. In each iteration of the main loop, we pair a \emph{donor column}, i.e., a column with positive surplus, with a \emph{receiver column}, i.e., a column with negative surplus, so as to enable the exchange/redistribution of tokens between the two columns. This process is repeated as long as there exists a column with negative surplus. Whenever a column becomes saturated, it is solved using a 1D reconfiguration algorithm. Finally, when the loop terminates, any unsolved column is guaranteed to have non-negative surplus and is therefore solved using the 1D algorithm. 

\subparagraph{Improved red-rec algorithm.}
We now describe our improved version of red-rec (see Alg.~\ref{alg:redrec}) whose implementation achieves greater operational and runtime performance.
This improved version retains most of the core features and ideas used in the earlier version~\cite{Cimring2023}, with two major improvements.

The first major improvement is a new approach for the computation of trajectories for tokens that are redistributed from a donor column to a receiver column.
Unlike the previous version of red-rec, where a brute-force approach was used, we cast this ``redistribution'' problem as a 1D problem and solve it using the exact 1D algorithm.
Consider a donor-receiver pair $(\mathcal{D}_i,\mathcal{R}_j)$, where we now use $\mathcal{D}_i$ to denote a donor column (at index $i$) and $\mathcal{R}_j$ to denote a receiver column (at index $j$).
First, we know, by design, that all columns between $i$ and $j$ must be solved (before $\mathcal{D}_i$ and $\mathcal{R}_j$ are paired together) and therefore the reservoir region between $i$ and $j$ is empty, allowing the free movement of tokens from donor to receiver.
Secondly, we can always assume, without loss of generality, that the reservoir region of the receiver $\mathcal{R}_j$ does not contain tokens before the transfer from donor column to receiver column begins; otherwise, since those tokens must eventually enter the target region, we can simply execute the minimum number of displacement operations needed to clear the reservoir.
Putting it all together, to create an instance of the 1D problem, we simply create a path consisting of the target region of $\mathcal{R}_j$ and the reservoir region(s) of $\mathcal{D}_i$, leaving the source and target vertices as they appear in their respective regions.
Since all tokens must traverse the same ``horizontal'' distance from the donor to the receiver, we can simply ignore this distance and apply their corresponding moves after solving the 1D problem. 

The second major improvement is a new strategy for pairing donors and receivers combined with the notion of delayed moves that guarantees that every token moves at most once.
Observe that when a donor does not have enough surplus to saturate a receiver, we do not have enough information yet to deduce the final positions of the redistributed tokens.
Therefore, instead of moving the tokens to arbitrary locations on the receiver, and then move them again when more tokens are received, we remember the moves by marking the corresponding tokens and delay moving them until we have enough information.
That is, we delay the moves until the receiver column is ready to be saturated by its current (final) donor; only then can we guarantee that the tokens on the receiver will never have to move again. 

\begin{algorithm}[H]
\caption{-- The red-rec algorithm 
}\label{alg:redrec}
\begin{algorithmic}[1]
\Require A grid graph, $G=(V,E)$, representing a static trap array; an initial configuration of atoms represented as a set of source vertices, $S \subseteq V(G)$; and a target configuration of atoms represented as a set of target vertices, $T \subseteq V(G)$ ($|S| \geq |T|$).
\For{each column $\mathcal{C}_i$, $0 \leq i < W$}
\State Compute surplus $\sigma_i = |S \cap V(\mathcal{C}_i)| - |T \cap V(\mathcal{C}_i)|$;
\If{$\sigma_i = 0$}
\State Solve column $\mathcal{C}_i$ using the exact 1D algorithm;
\EndIf
\EndFor
\While{there exists column with negative surplus}
\State Let $(\mathcal{C}_i,\mathcal{C}_j)$ denote a \textbf{best} donor-receiver pair;
\If{$|\sigma_i| - |\sigma_j| > 0$}
\State Reconfigure $|\sigma_j|$ tokens from $\mathcal{C}_i$ to $\mathcal{C}_j$;
\State Solve the receiver column $\mathcal{C}_j$ (exact 1D);
\ElsIf{$|\sigma_i| - |\sigma_j| = 0$}
\State Reconfigure tokens from $\mathcal{C}_i$ to $\mathcal{C}_j$;
\State Solve both $\mathcal{C}_i$ and $\mathcal{C}_j$ (exact 1D);
\Else 
\State Solve the donor column $C_i$ (exact 1D);
\State Mark excess tokens in $C_i$ for delayed moving to $\mathcal{C}_j$;
\EndIf
\State Update surpluses and pairs; 
\State Write delayed moves when the need arises;
\EndWhile
\For{each unsolved column $\mathcal{C}_i$, $0 \leq i < W$}
\State Solve column $\mathcal{C}_i$ using the exact 1D algorithm;
\EndFor
\end{algorithmic}
\end{algorithm}

More specifically, assuming without loss of generality that $i < j$, $\sigma_i > 0$, and $\sigma_j < 0$, we select a donor-receiver pair $(\mathcal{C}_i, \mathcal{C}_j)$ such that every column $\mathcal{C}_k$, $i < k < j$ has already been solved (and $\sigma_k = 0$), and such that the number of tokens that can be exchanged between the donor and receiver columns is maximized. If several such pairs exist, then we look to minimize the number of columns in between the donor and receiver columns to ``heuristically'' minimize the total number of displacement operations. If we still have several candidate pairs with a tie, we choose the pair that has a receiver that is closest to saturation. 

After a donor-receiver pair is selected, we either immediately execute the exchange of tokens between donor and receiver or mark appropriate tokens from the donor for delayed exchange to the receiver. In particular, one of the following three cases applies:

\begin{enumerate}
    \item Only the receiver column will have surplus zero, in which case the receiver column is solved using the exact 1D algorithm.
    \item Both the donor and receiver columns will have surplus zero, in which case we solve both columns using the exact 1D algorithm.
    \item Only the donor column will have surplus zero, in which case the donor column is solved using the exact 1D algorithm. In this case, we do not execute the moves that redistribute the tokens from donor to receiver immediately; instead we mark those tokens for delayed movement. Simply put, marked tokens no longer count towards the surplus of their original column and are considered to be part of the receiver column as far as computing surpluses is concerned. Once a receiver column becomes saturated, i.e., by being paired with enough donors, all moves corresponding to tokens marked for delayed movement into that column are executed.
\end{enumerate}

\subparagraph{Complexity.}
In terms of complexity, running the exact 1D algorithm on a column of the grid requires $\mathcal{O}(\sqrt{n})$-time to solve the assignment problem but potentially $\mathcal{O}(\sqrt{n} \cdot \sqrt{n}) = \mathcal{O}(n)$-time to produce the output list of moves, in the worst case. Hence, if all columns start with non-negative surplus, the red-rec algorithm solves the instance (and produces the output) in $\mathcal{O}(n\sqrt{n})$-time in the worst case. Since every iteration of the main loop reduces the number of unsolved columns by at least $1$, we can bound the number of iterations by the number of columns, i.e., $\sqrt{n}$. Without including the low-level details, we claim that every iteration can be implemented to run in $\mathcal{O}(n)$-time; note that there are at most $\sqrt{n}$ pairs to consider since a receiver column can only be paired with one of two columns, i.e., the donor is  the first positive-surplus column appearing either before or after the receiver. Putting it all together, the red-rec algorithm  solves the reconfiguration problem on a $\sqrt{n} \times \sqrt{n}$-grid in time $\mathcal{O}(n\sqrt{n})$.

\subsection{The red-rec algorithm (parallel implementation)}\label{sec-red-rec-parallel}

In this section, we describe how we parallelize our enhanced red-rec algorithm using GPUs.
A GPU can execute a very large number of threads concurrently.
These threads are grouped into thread blocks, where coordination across threads within the same block is efficient, whereas coordination across different blocks introduces greater overhead.
In our parallelization of the red-rec algorithm, we assign each thread block to reconfigure an entire column.
Threads in the same block work together to calculate the surplus of the column using a parallel reduction operation~\cite{wen2022programming}, and to reconfigure the column using our parallel exact 1D algorithm described in Sec.~\ref{sec-exact_1d-parallel}.
Different blocks reconfigure different columns.

To fully utilize the GPU hardware resources, we aim to execute as many thread blocks simultaneously as possible.
Hence, we must extract as many columns as possible that can be reconfigured simultaneously.
However, the sequential red-rec algorithm is designed to reconfigure the columns one after the other.
To extract columns that can be reconfigured in parallel, we employ two strategies. First, we reconfigure columns with positive surplus at the beginning along with columns with zero surplus. Second, we extract a partial order of the columns with negative surplus and group together those that can be reconfigured independently. The rest of this section describes the details of our parallel implementation.

\subparagraph{Reconfiguring columns with non-negative surplus.}
In the sequential red-rec algorithm, columns with zero surplus are reconfigured first because they are neither donors or receivers.
On the other hand, columns with positive surplus are reconfigured during the main loop of the algorithm or at the end, after they have donated to the receiver columns they are paired with.
In the parallel algorithm, to increase the number of columns that can be reconfigured simultaneously, we reconfigure both zero surplus and positive surplus columns from the outset.
All the columns are reconfigured in parallel by different thread blocks.

\subparagraph{Reconfiguring columns with negative surplus.}
Once the columns with non-negative surplus have been reconfigured, the columns with negative surplus remain.
In the sequential red-rec algorithm, the main loop of the algorithm repeatedly identifies the best donor-receiver pair, redistributes atoms across the pair, then reconfigures at least one of the columns in the pair.
The redistribution-reconfiguration cycle is repeated until all negative surplus columns have been saturated.
To parallelize such an inherently sequential algorithm, we observe that although identifying the pairs is a sequential process, actually solving the pairs via redistribution and reconfiguration can sometimes be done in parallel.
For example, consider the case where the first pair to be solved consists of columns \textit{a} and \textit{b}, and the second pair to be solved consists of columns \textit{c} and \textit{d}.
Although we cannot identify that $(c, d)$ is the second pair before identifying that $(a, b)$ is the first pair, we can actually solve $(a, b)$ and $(c, d)$ independently once they have been identified.
Hence, instead of identifying $(a, b)$, solving $(a, b)$, identifying $(c, d)$, then solving $(c, d)$, as the sequential algorithm does, we can identify $(a, b)$ then $(c, d)$ sequentially, and then solve them both in parallel.
Of course, if the first pair was $(a, b)$ and the second pair was $(a, c)$, then they cannot be solved in parallel because they share a column.
But still, if the third pair was $(d, e)$, then it can be solved in parallel with $(a, b)$ before $(a, c)$ because it is independent of both.

Based on this observation, our parallel red-rec algorithm reconfigures columns with negative surplus by sequentially identifying the donor-receiver pairs and updating their surpluses but without immediately resolving them, constructing a partial order of the pairs, traversing the partial order to construct groups of pairs that can be solved independently, then finally solving the pairs in the same group in parallel.
The process of identifying the groups of donor-receiver pairs to be executed in parallel is performed by a single thread block while the remaining thread blocks remain idle.
Once that thread block is done, it broadcasts the grouping to the other thread blocks.
The thread blocks then iterate through the groups, and for each group, solve the pairs in parallel, with each thread block solving a different pair.
Note that in both the sequential and parallel algorithms, the pairs are identified according to the same rules, the receivers are solved in the same way, and every token is guaranteed to move at most once.
The concept of delayed moves is also preserved. 

\subparagraph{Single kernel implementation.}
Since we evaluate our implementation on small problem sizes, the overhead of calling a GPU kernel can be significant.
To limit the impact of this overhead, we implement the entire parallel red-rec algorithm using a single GPU kernel.
Doing so also allows frequently used data to remain in the GPU's shared memory, avoiding the need to reload it from global memory across kernel calls.
To enable a single-kernel implementation, each thread block must have enough threads to map each thread to a trap within a specific column or map each thread to a column.
For this reason, we configure the kernel with $W$ thread blocks each having $\textsf{max}(H, W)$ threads, where $H$ and $W$ denote the height and width of the input graph, respectively.



\subsection{The bird algorithm}\label{sec-bird}

In the section, we describe the newly-proposed bird algorithm~(see Alg.~\ref{alg:bird}).
The bird algorithm is a heuristic algorithm that solves atom reconfiguration problems on grids.
It achieves greater operational performance than the red-rec algorithm~(Sec.~\ref{sec-red-rec}).
However, unlike red-rec, which can be adapted to handle arbitrary target configurations on grids, the bird algorithm requires target configurations to be centered within the grid, meaning the target configuration must form a subgrid centered inside the input grid.
We call it the bird algorithm because compared to the red-rec algorithm which locally inspects columns in pairs, the bird algorithm has a bird's-eye view of the grid for redistributing atoms. 

\begin{algorithm}[H]
\caption{-- The bird algorithm}\label{alg:bird}
\begin{algorithmic}[1]
\Require A grid graph, $G=(V,E)$, representing a static trap array; an initial configuration of atoms represented as a set of source vertices, $S \subseteq V(G)$; and a target configuration of atoms represented as a set of target vertices, $T \subseteq V(G)$ ($|S| \geq |T|$).
\State Compute the surplus for each column $\mathcal{C}_i$, $0 \leq i < W$, where surplus is $\sigma_i = |S \cap V(\mathcal{C}_i)| - |T \cap V(\mathcal{C}_i)|$;
\State Solve columns with non-negative surplus using the exact 1D algorithm;
\State Let $\mathcal{C}$ be the leftmost column with negative surplus ($\varnothing$ if none exist);
\While{$\mathcal{C} \neq \varnothing$}
\State Build the generalized 1D problem instance;
\State Identify the best atoms (and moves) to fill the target 
\State region of $\mathcal{C}$ using the generalized exact 1D algorithm; 
\State Execute the moves;
\State Set $\mathcal{C}$ to leftmost unsolved column ($\varnothing$ if none exist);
\EndWhile
\end{algorithmic}
\end{algorithm}

Similar to the red-rec algorithm, the bird algorithm starts by computing the surplus of each column, which can be zero, negative, or positive.
Any column with a positive or zero surplus is deemed self-sufficient and is solved using the exact 1D algorithm (see Sec.~\ref{sec-exact_1d}).
The moves required for each column are then saved and executed to modify the current input instance.
Once there are no more unsolved columns with a positive or zero surplus, the bird algorithm focuses on the remaining columns with negative surplus.
The key difference between the red-rec algorithm and the bird algorithm is in how these negative-surplus columns are handled.

\subparagraph{Drawback of the red-rec algorithm.}
The red-rec algorithm requires a receiver column to use all surplus tokens present in a donor column before receiving tokens from another donor column.
This requirement to exhaust surplus tokens in a column before using other tokens may result in moving distant tokens when there are better options available. 
For example, consider a case where a column $\mathcal{C}_1$ has two missing tokens, and is near two columns $\mathcal{C}_0$ and $\mathcal{C}_2$, each having two surplus tokens, one near the target region and one far away another at the extremity of the reservoir region.
In this case, the red-rec algorithm will match $\mathcal{C}_1$ with either $\mathcal{C}_0$ or $\mathcal{C}_2$, which results in moving one nearby token and one distant token.
However, it is clearly more efficient for $\mathcal{C}_1$ to receive one nearby token from $\mathcal{C}_0$ and one nearby token from $\mathcal{C}_2$.
\subparagraph{Handling negative surplus columns with the bird algorithm.}
To address this drawback, the bird algorithm adopts a many-to-one strategy while solving negative surplus columns instead of the one-to-one pairing strategy of red-rec.
That is, when solving a column, the bird algorithm considers tokens in all the other columns at once, not just a single column.
It does so by converting the many-to-one redistribution problem to a 1D problem where multiple tokens can occupy a single vertex.
In other words, tokens that are in different columns and have the same distance to the target region of the column being solved are logically treated as occupying the same vertex of that column.
The tokens do not physically occupy the same vertex, but rather, it is only a logical mapping for the purpose of formulating the problem in 1D. 
We solve this problem with our generalized exact 1D algorithm, which identifies the best tokens to move in our 1D array.
We then map these tokens back to their original positions, determine their locations in the grid, and execute the moves. 

More specifically, we map tokens in different columns to the column being solved as follows.
Assume that the algorithm is currently solving column $\mathcal{C}$.
Since our target area is centered, a token in the reservoir region of column $\mathcal{C'} \neq \mathcal{C}$ can be represented in any other column, with its location determined based on the distance between $\mathcal{C}$ and $\mathcal{C}'$, which we denote by $\textsf{dist}(\mathcal{C}, \mathcal{C}')$.
That is, assuming that vertices in a column are indexed from top to bottom starting from $0$, a token at vertex $h$ in $\mathcal{C}$ will be equivalent (in terms of distance with respect to $\mathcal{C}$'s target region) to an atom at vertex $h + \textsf{dist}(\mathcal{C}, \mathcal{C}')$ if it is in the top reservoir of $\mathcal{C}'$, or $h-\textsf{dist}(\mathcal{C}, \mathcal{C}')$ if it is in the bottom reservoir of $\mathcal{C}'$.
This follows from the fact that the shortest distance to the target region is the Manhattan distance, since we need to cover the horizontal and the vertical distances. 
It follows that tokens in a downward diagonal originating from a vertex in the top reservoir of a column $\mathcal{C}$ are all equivalent and will hold the same position on that specific vertex (in the generalized 1D instance).
Similarly, tokens in an upward diagonal originating from a vertex in the bottom reservoir of a column $\mathcal{C}$ are all equivalent and will hold the same position on that vertex.
If a token is picked from a vertex with several tokens, we use the token from the closest possible column as we are solving columns from left to right.
As such, we guarantee that tokens moving from one column to another will not encounter any obstructions, as any token along the way will be closer to the specific trap we are looking to fill and would have been chosen instead.

\subparagraph{Complexity.} 
Similarly to red-rec, running the (generalized) exact 1D algorithm on a column of the grid requires $\mathcal{O}(\sqrt{n} \cdot \sqrt{n}) = \mathcal{O}(n)$-time to solve the problem in the worst case. Hence, if all columns start with non-negative surplus, the bird algorithm solves the instance and produces the output in $\mathcal{O}(n\sqrt{n})$-time in the worst case. Since every iteration of the main loop reduces the number of unsolved columns by one, we can bound the number of iterations by the number of columns, i.e., $\sqrt{n}$. Every iteration consists of a grid traversal ($\mathcal{O}(n)$-time), followed by the construction of a (generalized) exact 1D instance ($\mathcal{O}(n)$-time), and concluded by a call to the generalized exact 1D algorithm ($\mathcal{O}(n)$-time). Putting it all together, the bird algorithm can solve the reconfiguration problem on a $\sqrt{n} \times \sqrt{n}$-grid in time   $\mathcal{O}(n\sqrt{n})$, matching the running time of red-rec asymptotically.

\subsection{The aro algorithm}\label{sec-aro}
The assignment-rerouting-ordering (aro) algorithm~\cite{ElSabeh2023} exploits exact and heuristic subroutines to solve atom reconfiguration problems on general graphs with a mixed objective function involving both the total number of displacement operations and the total number of displaced atoms. 
Compared to the red-rec and bird algorithms which are designed for grids, aro does not impose any restrictions on the input graph and the sets $S$ and $T$. It typically achieves better performance in terms of minimizing the total number of displacement operations and displaced atoms but requires a longer runtime. 
In this paper, we propose an improved version of the aro algorithm that decreases the running time of two of its subroutines.

\subparagraph{Original aro algorithm.}
The original aro algorithm proposed in Ref.~\cite{ElSabeh2023} consists of three main subroutines:
\begin{enumerate}
    \item The assignment subroutine computes a minimum-weight matching $M$ that saturates $T$ by running a matching algorithm on the bipartite graph $H$ constructed from the source vertices, target vertices, and the distances between them. As a result, every target vertex is paired with a source vertex while minimizing the sum of distances between the pairs. Using $M$, an initial path system $\mathcal{P}$ is constructed by computing an arbitrary shortest path in $G$ between every matched pair of vertices in $M$ using breadth-first search. Both $M$ and $\mathcal{P}$ are distance-minimizing, i.e., $w_H(M) = w_G(\mathcal{P})$. 
    \item The distance-preserving rerouting subroutine heuristically attempts to decrease the number of displaced tokens by modifying the paths in $\mathcal{P}$. The output of the subroutine is a new path system $\mathcal{P}'$ such that $w(\mathcal{P}') \leq w(\mathcal{P})$ and $\mathcal{P}'$ potentially displaces fewer atoms than $\mathcal{P}$. In the following, we assume that the subroutine replaces $\mathcal{P}$ with $\mathcal{P}'$, so that $\mathcal{P}$ from here on  refers to the updated path system. 
    \item The ordering subroutine computes a partial ordering of the paths to ensure that executing the associated moves in this order guarantees that each token moves at most once while avoiding collisions.
    The reason why it is possible for the ordering subroutine to compute a partial ordering is not immediately obvious. 
    In fact, not all path systems admit such an ordering.
    However, one of our key results presented in Ref.~\cite{ElSabeh2023} states that any (valid) path system $\mathcal{P}$ can be transformed in polynomial time into a (valid) path system $\mathcal{P}'$ such that $w(\mathcal{P}') \leq w(\mathcal{P})$ and the graph induced on the edges of $\mathcal{P}'$, which we denote by $G_{\mathcal{P}'}$, is a forest, i.e., a graph with no cycles (we call such a path system a \emph{cycle-free} path system).
\end{enumerate}
The original aro algorithm has two major bottlenecks:
\begin{enumerate}
    \item In the assignment subroutine, the basic implementation of the Hungarian method~\cite{edmonds1972theoretical} runs in time $\mathcal{O}(n^3)$ to compute a distance-minimizing matching between sources and targets.
    This implementation comes with the additional requirement that, before applying the Hungarian method, we must construct an edge-weighted bipartite graph $H = (L \cup R, E')$ (as described in Sec.~\ref{sec-matching-flows-systems}), with $L = S$, $R = T$, and where the weight of an edge in $H$ is set to the distance between the corresponding vertices in $G$.
    \item In the ordering subroutine, the cycle-breaking procedure constitutes a main bottleneck since, as shown in Ref.~\cite{ElSabeh2023}, the bulk of the work in the original aro algorithm is the design of an efficient procedure to transform any path system into a cycle-free path system without increasing the weight.
\end{enumerate}

Our proposed improved aro algorithm aims to address these two bottlenecks.

\subparagraph{Improved aro algorithm.}
The improved aro algorithm~(see Alg.~\ref{alg:aro}) begins by computing shortest paths between vertices in $S$ and $T$, similar to the original aro algorithm.
It then computes a distance-minimizing matching between sources and targets to generate a path system $\mathcal{P}$.
However, unlike the original aro algorithm, which uses the Hungarian method, the improved aro algorithm reduces the problem of finding a distance-minimizing matching to the minimum-cost maximum-flow (MCMF) problem when the graphs are unweighted. MCMF is more efficient than the Hungarian method for graphs where the number of edges is linear in the number of vertices, such as the planar graphs we are interested in. 

If the graph induced by the edges of $\mathcal{P}$, $G_{\mathcal{P}} = (V, \mathcal{E})$, is a forest, it immediately generates a partially-ordered list of displacement operations with the required properties and terminate (see Ref.~\cite{ElSabeh2023} for a proof of existence of such an ordering).
Otherwise, it constructs a valid distance-minimizing path system that is cycle-free using an improved cycle-breaking procedure.

\begin{algorithm}[H]
\caption{-- The aro algorithm}\label{alg:aro}
\begin{algorithmic}[1]
\Require A tuple $(G,S,T)$ where $G=(V,E)$ is an $n$-vertex grid graph, $S \subseteq V(G)$ is the source configuration of tokens, $T \subseteq V(G)$ is the target configuration of tokens, and $|S| \geq |T|$;

\State For every vertex in $u \in T$, compute a shortest path from $u$ to every vertex $v \in S$ using breadth-first search in $G$~($\mathcal{O}(n^2)$);

\State Compute a distance-minimizing matching $M$ followed by a distance-minimizing path system $\mathcal{P}$ (we assume $w_G(\mathcal{P})$ is finite) using the MCMF solver~($\mathcal{O}(n^2)$); 

\State Apply the (distance-preserving) rerouting heuristic to reduce the number of displaced tokens~($\mathcal{O}(n^3)$);

\While{$G_{\mathcal{P}}$ is not a forest~($\mathcal{O}(n)$)}
\State Let $\mathcal{E} = E(G_{\mathcal{P}})$ (edges of the path system);
\For{each edge $e \in \mathcal{E}$}\
\State Let $G' = (V, \mathcal{E} \setminus \{e\})$;
\State Compute shortest paths between $S$ and $T$ in $G'$;
\State Compute $\mathcal{P}'$ for $(G', S, T)$~($\mathcal{O}(n^2)$);
\If{$w_{G'}(\mathcal{P}') \leq w_G(\mathcal{P})$}
\State Let $\mathcal{P} = \mathcal{P}'$;
\State Let $G = G'$;
\State \textbf{break};
\EndIf
\EndFor
\EndWhile
\State Generate partially-ordered \textsf{list} of moves~($\mathcal{O}(n^2)$);
\State \Return \textsf{list};

\end{algorithmic}
\end{algorithm}

\subparagraph{MCMF-based assignment routine.}
The improved aro algorithm reduces the problem of finding a distance-minimizing matching to the MCMF problem, rather than using the Hungarian method, when the graphs are unweighted.
In the standard formulation of the MCMF problem, the aim is to maximize a flow sent from a source to a sink vertex while minimizing a predetermined notion of cost, with the edges having costs and capacities.
We let $H$ be a copy of $G$ where we then replace every edge by two directed edges, one in each direction, each of which having infinite capacity and unit cost.
We add a vertex $s^\star$ to $H$ and add directed edges from $s^\star$ to every vertex of $S$. Similarly, we add a vertex $t^\star$ to $H$ and add directed edges from every vertex in $T$ to $t^\star$.
Edges incident to $s^\star$ or $t^\star$ have unit capacity and unit cost. It is known that a distance-minimizing matching that saturates $T$ corresponds to a minimum-cost flow of value $|T|$ from $s^\star$ to~$t^\star$~\cite{tarjan1997dynamic}. 

Our MCMF solver is similar in essence to the Edmonds-Karp algorithm, an implementation of the Ford-Fulkerson method to compute maximum flows in a network~\cite{edmonds1972theoretical, ford1956maximal}.
At every step, a shortest (augmenting) path between $s^\star$ and~$t^\star$ is computed, and flow is sent along this path, while keeping track of the directions along which the flow units are being sent in.
The computation of the shortest path can be done in time $\mathcal{O}(n + m)$ using breadth-first search, because our graphs are unweighted.
The graph $H$ can be constructed in time $\mathcal{O}(n + m)$.
Since computing a matching will require computing $\mathcal{O}(n)$ shortest paths, and given that $m \in \mathcal{O}(n)$ in planar graphs, our MCMF solver runs in time $\mathcal{O}(n^2)$ (in contrast to the running time of $\mathcal{O}(nm\log n)$ for the algorithm that solves the problem for unweighted arbitrary graphs on $n$ vertices and $m$ edges \cite{tarjan1997dynamic}).

\subparagraph{Improved cycle-breaking procedure.}
The other, more substantial improvement that we propose to the aro algorithm is an improved cycle-breaking procedure.
When $G_{\mathcal{P}}$ is not a forest, the main result of Ref.~\cite{ElSabeh2023} guarantees the existence of at least one valid distance-minimizing path system $\mathcal{P}^{\star}$, such that $G_{\mathcal{P}^\star}$ is a forest, $G_{\mathcal{P}^\star}$ uses a proper subset of the edges of $G_{\mathcal{P}}$ and $w(\mathcal{P}^\star) \leq w(\mathcal{P})$.
This existence in turn implies that there exists at least one edge $e$ in $\mathcal{E}$ that does not need to be present in a valid distance-minimizing path system.
We say that $e$ is \emph{safe}, because deleting it does not affect the existence of a valid distance-minimizing path system.
To determine whether an edge is safe to delete, we delete the edge and compute a distance-minimizing path system $\mathcal{P}'$ in the resulting graph with one less edge $G' = (V, \mathcal{E} \setminus \{e\})$.
If $w_{G'}(\mathcal{P}') \leq w_G(\mathcal{P})$ then $e$ is safe and we retain the edge deletion and replace $G$ and $\mathcal{P}$ by $G'$ and $\mathcal{P}'$, respectively, and repeat the process as long as $G_{\mathcal{P}}$ is not a forest.
Termination is again guaranteed by Theorem 1 in~\cite{ElSabeh2023}.
After finitely many edge deletions, we obtain the path system $\mathcal{P}^\star$, as desired.

Each iteration of the improved cycle-breaking procedure reduces the number of edges by one.
Moreover, the number of edges in a planar graph is $\mathcal{O}(n)$.
Hence, the main loop in the algorithm iterates $\mathcal{O}(n)$ times.
In the worst case, at any given point, exactly one edge of $\mathcal{E}$ is safe to delete and that edge appears at the last iteration of the for loop.
As $|\mathcal{E}| = \mathcal{O}(n)$, a single iteration of the main loop attempts to delete $\mathcal{O}(n)$ edges, one edge at a time, and for every attempt, a distance-minimizing path system is computed.
This implies that $\mathcal{O}(n^2)$ distance-minimizing path systems are computed in the worst case.
Given that a distance-minimizing path system can be computed in time $\mathcal{O}(n^2)$ using the MCMF approach, the revised version of the aro algorithm runs in time $\mathcal{O}(n^4)$ while our previous, more general, implementation runs in time $\mathcal{O}(n^8)$.
Note, however, that the revised version of the aro algorithm does not render the original version obsolete, as the revised version does not support weighted graphs and is tailored for grid graphs only.

\section{Batching routine}\label{sec-batching}

Batching groups together moves that can be executed simultaneously, which improves the performance of executing the list of moves generated by the reconfiguration algorithms.
In prior work, the 1D and red-rec algorithms generated batched moves implicitly, whereas the aro algorithm did not generate batched moves.
In this paper, we propose a general batching routine that works for any of our proposed reconfiguration algorithms, including the exact 1D, red-rec, bird, and aro algorithms.
In this section, we describe this batching routine in its most general form, i.e., by ignoring restrictions associated with input parameters and other restrictions imposed by the control system.
We then discuss how to adapt the routine to handle specific constraints that may be imposed by the system or the problem input parameters.

\subparagraph{Input to the batching routine.}
Recall that once we solve a reconfiguration problem, the output is an ordered list of unobstructed moves represented by a path system $\mathcal{P} = \{P_1, P_2, \ldots, P_k\}$, where each $P_i \in \mathcal{P}$ for $i \in \{1, \ldots, k\}$ is a  path from $v^i_0$ to $v^i_{\ell_i}$, which we denote by $\{v^i_{0}, v^i_1, v^i_2, \ldots, v^i_{\ell_i}\}$. Hence, an elementary displacement operation consists of moving a token from some vertex $v^i_j$ to vertex $v^i_{j+1}$ along the edge $\{v^i_j,v^i_{j+1}\}$ connecting them in the path $P_i$. Instead of an ordered list of moves $\mathcal{P}$, we assume that we are also given, in addition to an unordered set $\mathcal{P}$, a directed acyclic graph $\Gamma$ representing the dependencies between the moves in $\mathcal{P}$. In other words, the vertex set of $\Gamma$ contains one vertex $p_i$ for each $P_i \in \mathcal{P}$ and a directed edge from $p_i$ to $p_j$ exists if and only if $P_i$ must be executed before $P_j$ to avoid obstructions. Requiring $\Gamma$ to encode dependencies instead of simply ordering $\mathcal{P}$ allows for much greater freedom in the construction of batches. 

For most cases, $\Gamma$ can be computed while the reconfiguration problem is being solved, regardless of the algorithm, and we therefore exclude it from our runtime analysis. For the exact 1D algorithm, $\Gamma$ is easily inferable since we can partition the moves into two disjoint sets of left moves and right moves (there are no edges in $\Gamma$ between the two sets), as we describe in Sec.~\ref{sec-exact_1d}. The left (right) moves can be ordered from left (right) to right (left) using target vertices, and only moves that intersect along the path need to be connected in $\Gamma$. For the red-rec and bird algorithms, we follow the same strategy as in the 1D case when solving individual columns. We use an almost identical strategy when distributing atoms between columns. To keep the implementation simple, we force all the moves required for solving column $\mathcal{C}_{i}$ to occur after the moves required for column $\mathcal{C}_{j}$ whenever $\mathcal{C}_{i}$ is solved after $\mathcal{C}_{j}$ and solving $\mathcal{C}_{i}$ requires at least one vertex that was used for solving $\mathcal{C}_{j}$. For aro, we implicitly construct the graph $\Gamma$. Now, armed with $\mathcal{P}$ and $\Gamma$, we are  ready to describe the batching routine. 

\subparagraph{Batching algorithm.}
The batching algorithm~(see Alg.~\ref{alg:batching}) constructs batches iteratively, saving them inside an ordered list $\mathcal{B}$. After every iteration, it reduces the size of $\mathcal{P}$ by deleting at least one edge, guaranteeing termination after at most $n^2$ iterations. In each iteration, we initialize an empty batch, $B = \emptyset$, to which we will add at most one elementary displacement operation (or equivalently one edge) from each $P \in \mathcal{P}$. Specifically, we consider each path in $\mathcal{P}$ whose vertex in $\Gamma$ has no incoming edges, i.e., in-degree zero vertex in $\Gamma$. Let $\mathcal{Q} = \{ P_1, P_2, \ldots, P_q\}$ denote those paths. For each $P_i \in \mathcal{Q}$, we check whether $\{v^i_0, v^i_1\}$ intersects with any set (edge) already added to $B$ (note that they can only intersect on the second vertex). 
If no intersection is found, we proceed as follows. We let $B = B \cup \{v^i_0, v^i_1\}$, we update $P_i$ by deleting $v^i_0$ and the edge connecting it to $v^i_1$ (we denote this operation by $P_i = P_i - \{v^i_0\}$ which we also assume renumbers the vertices starting from $0$). If $P_i$ now consists of a single vertex, we delete $P_i$ from $\mathcal{P}$ and delete the vertex corresponding to $P_i$, i.e., $p_i$, from $\Gamma$ (edges incident to $p_i$ are also deleted, i.e., $\Gamma = \Gamma - \{p_i\}$); intersecting edges are ignored. Once no new edges of $\mathcal{Q}$ can be added to $B$, we have an inclusion-wise maximal set of elementary displacement operations that can be executed simultaneously, which we add to $\mathcal{B}$, i.e., $\mathcal{B} = \mathcal{B} \cup B$. The algorithm terminates once all the edges of the path system have been added to some batch, i.e., when the path system (and the  directed acyclic graph) become empty. 

\begin{algorithm}[H]
\caption{-- The batching routine}\label{alg:batching}
\begin{algorithmic}[1]
\Require A tuple $(G,S,T,\mathcal{P},\Gamma)$ where $G=(V,E)$ is an $n$-vertex graph, $S \subseteq V(G)$ is the source configuration of tokens, $T \subseteq V(G)$ is the target configuration of tokens, $|S| \geq |T|$, $\mathcal{P}$ is a distance-minimizing path system, and $\Gamma$ is a directed acyclic graph representing the unobstructed partial ordering of the moves in $\mathcal{P}$;
\State Let $\mathcal{B} = \emptyset$;
\While{$\mathcal{P}$ contains at least one edge}
\State Compute $\mathcal{Q} = \{ P_1, P_2, \ldots, P_q\}$ the set of paths having 
\State in-degree $0$ in $\Gamma$;
\State Let $B = \emptyset$;
\For{$P_i \in \mathcal{Q}$}
\State Let $e = \{v^i_0, v^i_1\}$;
\If{$\forall e' \in B$ we have $e \cap e' = \emptyset$}
\State Let $B = B \cup e$;
\State Let $P_i$ = $P_i - \{v^i_0\}$ (renumber);
\If{$|V(P_i)| = 1$}
\State Let $\mathcal{P} = \mathcal{P} \setminus P_i$; 
\State Let $\Gamma = \Gamma - \{p_i\}$; 
\EndIf
\EndIf
\EndFor
\State Let $\mathcal{B} = \mathcal{B} \cup B$;
\EndWhile
\State \Return $\mathcal{B}$;
\end{algorithmic}
\end{algorithm}

\subparagraph{Constrained batching.}
Algorithm~\ref{alg:batching} assumes no restrictions are imposed by the control system. However, most control systems are limited in the types of simultaneous moves that they can perform. For example, in the case of 2D grids, one might want to allow batches consisting only of atoms appearing contiguously along the same column (or row). Another constraint is that of directionality, i.e., a batch should consist of atoms moving in the same direction in the grid (up, down, left, or right). As long as all constraints are what we call local, we can adapt our algorithm by augmenting the input tuple  $(G,S,T,\mathcal{P},\Gamma)$ with one extra parameter $\Lambda$, where $\Lambda$ is a list of constraints we want to satisfy. We say that a constraint is \emph{local} whenever it suffices to check whether it holds for (at most) every pair of elements in a batch. Then, as long as all constraints are local, the only modification required in Alg.~\ref{alg:batching} is on Line~8, where we have to also verify that all constraints in $\Lambda$ are satisfied. 
The current batching routine deployed on our system only allows batches consisting of tokens in the same column (row) and moving in the same direction 

\subparagraph{Complexity}
When $\Gamma$ consists of a directed path, the batching routine runs in worst-case $\mathcal{O}(n^2)$-time, since the path system could consist of $\mathcal{O}(n^2)$ edges in the worst case (resulting in $\mathcal{O}(n^2)$ batches).
Moreover, for ``reasonable'' sets of local constraints (such as ours), the $\mathcal{O}(n^2)$-time bound holds regardless of the structure of $\Gamma$. 

\section{Quantifying performance}\label{sec-operational-benchmarking}

To justify their practical implementation, we evaluate our algorithms by measuring their runtime and operational performance.

\subsection{Performance evaluation methodology}

Runtime performance is quantified by the average time required to solve a specific reconfiguration problem, and its scaling with the problem size. It depends on the algorithm, its implementation, and the specifications of the processor, whether a CPU (AMD Ryzen Threadripper 2950X) or a GPU (Nvidia Quadro RTX 4000). 
Operational performance is quantified by the mean success probability of solving a specific reconfiguration problem when randomly sampling over the initial configuration of atoms and loss processes. It depends on the reconfiguration algorithm and batching routine, as well as the loss parameters. It is computed numerically using realistic physical parameters, following the approach outlined in our previous works~\cite{Cimring2023, ElSabeh2023}.

For trap arrays with 1D geometries, we solve the problem of preparing a center-compact chain of $N_a^T$ atoms in a chain of $N_t = N_a^T / \eta$ traps, where $\eta = 0.5$. 
For trap arrays with 2D geometries, we solve the problem of preparing a center-compact configuration of $N_a^T$ atoms in a rectangular grid of $N_t = N_{t_x} \times N_{t_y}$ traps with $N_{t_x}=\sqrt{N_a^T}$. Unless specified otherwise, when $N_{t_y}$ is fixed, we typically choose $N_{t_y}=N_{t_x}/\epsilon$ with $\epsilon = 0.6$. This number of traps is chosen to achieve a baseline success probability $\bar{p}_0 = 0.5$. The baseline success probability is the mean success probability obtained in the absence of loss, determined solely by the number of atoms loaded in the initial configuration~\cite{Cimring2023}. 

We solve each problem over a thousand problem instances. For each instance, we randomly sample the initial configuration of $N_a^0=\epsilon N_t$ atoms, as well as the loss realizations.  We choose a trapping lifetime of $\tau=60~\text{s}$, and a loss probability of $p_\nu=0.985$ and $p_\alpha=0.985$ for displacement and transfer operations, respectively. 
When using the GPU, we operate in persistence mode to keep it fully powered at all times. Otherwise, we typically observe that runtime performance improves over the first few repetitions as the GPU resources become fully active.

\subsection{Runtime performance}

\begin{figure}[t!]
\includegraphics[]{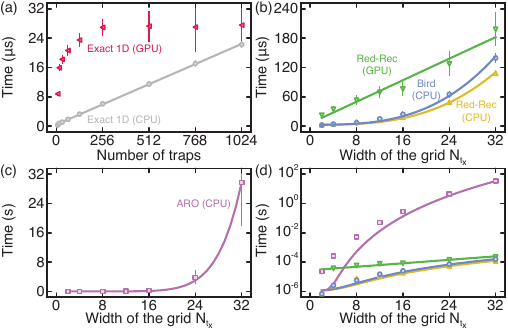}
\caption{
\label{fig-runtime}
\textbf{Runtime performance.}
(a)~Measured runtime for solving reconfiguration problems on chains using the serial implementation of the exact 1D algorithm on a CPU (gray disks) and its parallel implementation on a GPU (red triangles). The gray line is a linear fit to the CPU data. 
(b)~Measured runtime for solving reconfiguration problems on grids using the serial implementation of the red-rec algorithm on a CPU (yellow triangles), its parallel implementation on a GPU (green inverted triangles), and the bird algorithm on a CPU (blue disks). Solid lines are cubic fits for red-rec (CPU) and bird, and linear fits for red-rec (GPU).
(c)~Measured runtime for solving reconfiguration problems on grids using the aro algorithms. The purple line is a quartic fit to the data.
(d)~Same data as in (b) and (c) represented on a semi-log scale. 
}
\end{figure}

We first analyze the runtime performance of the exact 1D algorithm, considering both its serial implementation on a CPU~(see Sec.~\ref{sec-exact_1d}) and its parallel implementation on a GPU~(see Sec.~\ref{sec-exact_1d-parallel}).
The serial implementation is fast, solving a chain with $N_t=1024$ traps in $22(1)~\mu\text{s}$.
Its runtime scales linearly with the number of atoms (see gray line in Fig.~\ref{fig-runtime}a).
In contrast, the parallel implementation has a finite initialization time $7~\mu\text{s}$ for starting parallel GPU kernels. 
Its runtime is upper bounded by a constant factor, and it achieves better performance for small problem sizes.
Additionally, our implementation is restricted to chains of no more than 1024 traps due to the limitations in the number of threads per block on our GPU.
These results indicate that the serial implementation outperforms the parallel implementation when solving 1D problems. However, this conclusion may change with future hardware advancements, as GPUs continue to evolve rapidly.

We then analyze the runtime performance of the red-rec algorithm, considering both its serial implementation on a CPU~(see Sec.~\ref{sec-red-rec}) and its parallel implementation~on a GPU~(see Sec.~\ref{sec-red-rec-parallel}).
The serial implementation is fast, solving a grid of $N_a^T=32^2$ atoms in $106(6)~\mu\text{s}$. As expected, its runtime scales as $\mathcal{O}(N_t^{3/2}=N_{t_x}^3)$~(see yellow line in Fig.~\ref{fig-runtime}b,~d). The parallel implementation exhibits a finite initialization time of $7~\mu\text{s}$ and approximately linear scaling with the nearest power of two
of $N_{t_x}$ (see green line in Fig.~\ref{fig-runtime}b). 
For the small problem sizes considered here, the parallel implementation is slightly slower than the serial implementation, solving a grid of $N_a^T=32^2$ atoms in $197(35)~\mu\text{s}$. However, by extrapolating the fits, we predict a performance crossover at $N_{t_x}\approx42$.

We further analyze the runtime performance of the bird algorithm. Compared to red-rec, bird is relatively slower but still very fast, solving a grid of $N_a^T=32^2$ atoms in $138(10)~\mu\text{s}$. Its runtime scaling is the same as red-rec, $\mathcal{O}(N_t^{3/2}=N_{t_x}^3)$~(see blue line in Fig.~\ref{fig-runtime}b,~d). Given that bird achieves a slightly higher mean success probability than red-rec (see Sec.~\ref{sec-op-performance}), the slight increase in runtime may be justified for real-time operations.

We finally analyze the runtime performance of the aro algorithm~(see Sec.~\ref{sec-aro}). We have significantly improved its implementation since our previous work~\cite{ElSabeh2023}, reducing its worst-case time complexity from $\mathcal{O}(N_t^8)$ to $\mathcal{O}(N_t^4)$ (see purple curve in Fig.~\ref{fig-runtime}c,~d). Still, aro is significantly slower than both red-rec and bird, taking $30(12)~\text{s}$ to solve a grid of $N_a^T=32^2$ atoms. Although such a runtime prevents real-time operations in practical settings, our improved implementation remains valuable for cross-benchmarking the performance of other algorithms, including those designed to solve atom reconfiguration problems on arbitrary graphs.

\subsection{Operational performance}\label{sec-op-performance}

\begin{figure}[t!]
\includegraphics[]{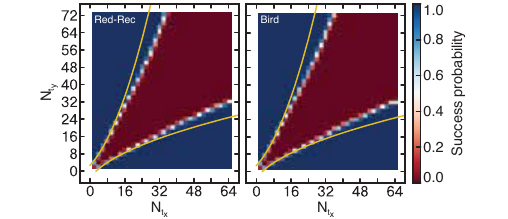}
\caption{
\label{fig-surface}
\textbf{Operational performance.}
Mean success probability for preparing a center-compact configuration of $N_{t_x}$ atoms in a rectangular grid of $N_{t_x} \times N_{t_y}$ traps using red-rec (left) and bird (right) with batching. Both algorithms outperform the deprecated version of red-rec, with the transition curve at $\bar{p} = 0.5$ outlined in orange.
}
\end{figure}

Having quantified the runtime performance of the various algorithms, we now focus on quantifying their operational performance.
We compute the mean success probability for various problem sizes (Fig.~\ref{fig-surface}).
The surface plot exhibits two distinct regions: one representing near-certain failure, with $\bar{p} \leq 0.02$, and the other representing near-certain success, with $\bar{p} \geq 0.98$.
These two regions are separated by a sharp transition region centered around $\bar{p} = 0.5$.
This transition region scales approximately as $N_{t_y}\sim N_{t_x}^{\alpha}$, with $\alpha=1.077$ for red-rec and $\alpha=1.047$ for bird.
These numbers indicate a need for $N_t\sim (N_a^T)^{(1+\alpha)/2}$ traps to prepare an array of $N_a^T$ atoms.
Both the bird and red-rec algorithms outperform our previous version of red-rec (orange curves in Fig~\ref{fig-surface}). 

\begin{figure}[th]
\includegraphics[]{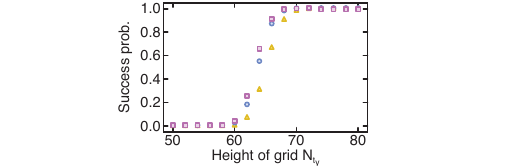}
\caption{
\label{fig-operational-performance}
\textbf{Operational performance.} 
Mean success probability for preparing a center-compact configuration of $N_a^T=32^2$ atoms using red-rec~(yellow triangles), bird (blue disks), and aro (purple square) without batching. The reconfiguration problem is a in a rectangular grid of $N_t=32\times N_{t_y}$ traps.
}
\end{figure}

We further compute the mean success probability for preparing a configuration of $N_a^T=32^2=1024$ atoms in an array of $N_t=32\times N_{t_y}$ traps~(Fig.~\ref{fig-operational-performance}). 
We observe that aro outperforms bird, which in turn outperforms red-rec. For example, for $N_{t_y}=64$, we find that $\bar{p}=0.30, 0.54, 0.65$ for red-rec, bird, and aro, respectively.

\begin{figure}[t!]
\includegraphics[]{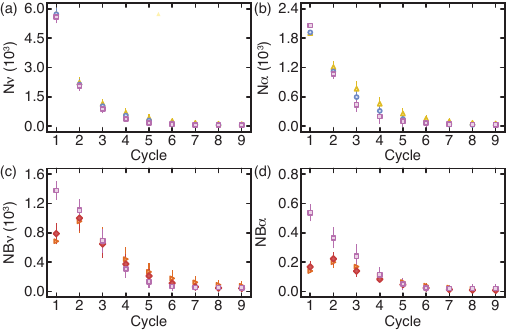}
\caption{
\label{fig-ops-per-cycle}
\textbf{Control operations.}
(a-b)~Mean number of displacement and transfer operations when solving a typical problem using red-rec (yellow triangles) and bird (blue disks) without batching.
(c-d)~Number of batched displacement and transfer operations when solving the same problem using red-rec (orange triangles) and bird (red diamonds) with batching.
}
\end{figure}
\begin{figure}[t!]
\includegraphics[]{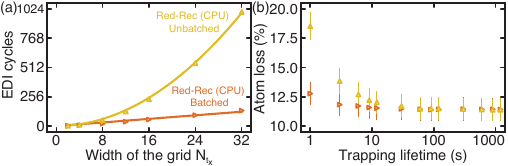}
\caption{
\label{fig-batching}
\textbf{Batching routine.} 
(a)~Distribution of the number of EDI cycles executed using the red-rec algorithm with the batching routine (blue disks) or without it (yellow squares). The reconfiguration problem is a center-compact configuration of $N_a^T=32^2$ atoms in a grid of $N_t=32\times64$ traps with a loading efficiency of $\epsilon=0.6$.
(b)~Mean number of EDI cycles for preparing a configuraiton of $N_a^T=N_{t_x}^2$ atoms in a rectangular grid of $N_t=N_{t_x}\times N_{t_y}$ traps with $N_{t_y}=2N_{t_x}$. The number grows quadratically in the absence of batching, as opposed to linearly in the presence of batching.
(c)~Percentage of atom loss for different trapping lifetimes. Batching becomes progressively less advantageous as the trapping lifetime approaches infinity.
(d)~Ratio of EDI cycles executed without batching to those executed with batching. The relative gain in performance increases linearly with the width of the grid.
}
\end{figure}

To better understand the reason for bird outperforming red-rec, we compute the number of displacement operations, $N_\nu$, and transfer operations, $N_\alpha$, over multiple reconfiguration cycles in the presence of loss.
We recall that aro exactly minimizes the number of displacement operations, albeit at the cost of a slight increase in the number of displaced atoms, and thus the number of transfer operations~\cite{ElSabeh2023}.
Without batching (Fig.~\ref{fig-ops-per-cycle}a,~b), bird performs fewer transfer and displacement operations than red-rec across all reconfiguration cycles, except for the first cycle, where the numbers are approximately the same.
With batching (Fig.~\ref{fig-ops-per-cycle}c,~d), the number of displacement and transfer operations is conserved, but the total execution time is reduced due to the batching of sequences of displacement operations, $\text{NB}_\nu$, and transfer operations, $\text{NB}_\alpha$.
We observe that red-rec has slightly fewer batched sequences of operations in the first two cycles, but bird performs better in the subsequent cycles.
However, because of the relatively long trapping lifetime ($\tau=60~\text{s}$), the relative decrease in the execution time resulting from batching does not translate in a significant increase in performance (see Fig.~\ref{fig-batching}b), at least not sufficient to outperform aro, which benefits less from batching.
These results outline an approach to construct a hybrid algorithm exploiting the red-rec algorithm in the first two reconfiguration cycles, and bird afterward.

\section{Conclusion}
In conclusion, we presented new and enhanced versions of atom reconfiguration algorithms. These algorithms, along with their efficient implementations, demonstrate improved runtime and operational performance. We showed that sequential implementations of the exact 1D and red-rec algorithms on CPUs outperform parallel implementations on GPUs for small problem sizes, but that GPUs might be justified for preparing grids of more than $N_a^T=42^2=1764$ atoms. We showed that bird, which was specifically designed as a generalization of red-rec for preparing a center-compact configuration of atoms, outperforms red-rec, at the cost of a slight increase in computational runtime. We further showed that the aro algorithm, which is valid for any arbitrary graph, still outperforms both red-rec and bird, but that its runtime is too slow for real-time applications, restricting its use to benchmarking applications. We finally showed that the batching routine decreases the total number of control sequences, and thus the total execution time, but that the advantage it provides is not significant as the trapping lifetime increases beyond $60~\text{s}$.

Our results highlight the typical trade-off between runtime performance and operational performance, and the importance of optimizing and benchmarking implementations on processors. They further establish formal groundwork for the development of improved atom reconfiguration algorithms and their deployment in practical settings. These algorithms can readily be used to solve atom reconfiguration problems and assess the performance of new algorithms.

Beyond these findings, future work will focus on integrating these algorithms into circuit compilation techniques for quantum processors with dynamic connectivity graphs, combining displacement operations with unitary control to realize digital quantum circuits. On the theoretical front, open questions remain regarding the development of a general theory of batching algorithms for reconfiguration problems defined on arbitrary graphs.

Further improvements can also be made to the bird algorithm to speed up its implementation. Currently, for each column with negative surplus, the entire grid is traversed to construct a generalized exact 1D instance from scratch. A more efficient approach would be to traverse the grid only once after resolving columns with zero or positive surplus and represent all atoms in the reservoir regions as diagonals in a dedicated data structure, enabling fast updates after each solved column. Another possible optimization is to transition from a many-to-one approach to a $2d$-to-one approach, sampling atoms only from columns within a fixed distance $d$ on either side. While this should not significantly impact operational performance, since columns are typically saturated by nearby atoms rather than those from opposite ends of the grid, it could substantially reduce computation time. However, ensuring correctness and termination requires choosing $d$ appropriately to guarantee sufficient available atoms.

Our source code is available for non-commercial use in a public repository.

\section{Acknowledgment}
This research was funded thanks in part to CFREF.

\clearpage


%

\end{document}